\documentclass[aps,prl,twocolumn,showpacs,superscriptaddress,groupedaddress]{revtex4}  
\usepackage{graphicx}  
\usepackage{subfigure}
\usepackage{caption}
\usepackage{bm,amsmath,amssymb,wasysym,amsfonts,dcolumn}
\usepackage[colorlinks=true, urlcolor=blue, linkcolor=blue, citecolor=blue, pdfborder={0 0 0}]{hyperref}
\usepackage[usenames,dvipsnames]{xcolor}
\usepackage{cleveref}
\usepackage{dcolumn}   
\usepackage{bm}        
\usepackage{amssymb}   
\usepackage{xcolor}
\usepackage{gensymb}
\usepackage{float}
\usepackage[T1]{fontenc}

\begin{document}



\title{Dynamic Response of Glassy Dispersions in a Nematic Liquid Crystal.}

  \author{N.Katyan}
\affiliation{School of Physics and Astronomy, University of Edinburgh, EH9 3FD, Edinburgh, U.K.}   


  \author{T.A.Wood}
\affiliation{School of Physics and Astronomy, University of Edinburgh, EH9 3FD, Edinburgh, U.K.}

\date{\today}

\begin{abstract}
We present the first systematic measurements of the dynamic response of a glassy soft-solid formed on dispersing colloids in a thermotropic nematic liquid crystal.  Through applying a simple theoretical model, we reveal that the nematic elasticity dominates at low frequencies (longest timescales) and prevents flow.   The Ericksen number $E_R$, determines a critical frequency $\omega_c$, above which the viscous component $G'' \propto \omega ^{1/2}$.  We find that the value of tan$\delta$ is independent of the colloidal volume fraction ($\phi$), and is determined by the ratio of lengthscales associated with the defect-line density and the size of colloid-free nematic regions trapped within the composite. 
\end{abstract}

\maketitle

Physical stability, preventing sedimentation or creaming, is key to the shelf-life of formulations and composites.  Stability can be assured if the storage modulus $G'$, describing the elasticity, is higher than the loss modulus $G''$, describing the viscous properties of the fluid.   Non-equilibrium structures such as gels and glasses allow dispersions to be distributed homogeneously for the `shelf-life'.  The microstructure can rearrange if the interaction energy $U\sim k_BT$. This is common for colloid-polymer mixtures and renders them sensitive to phase separation \cite{bergenholtz}. Twenty years ago, Poulin \textit{et al.}, dispersed colloids in a nematic liquid crystal and found $U>100k_BT$ \cite{Poulin1997}.   A novel defect-stabilised gel with $G' >10^3$Pa, is formed when colloids are dispersed at volume fractions $\phi > 0.18$ \cite{wood2011}.   For the first time, we report on the dynamics and yielding behaviour of these composites and provide a theoretical description of our experimental results.

Rheological spectra, obtained by applying small-angle oscillatory shear, offer a powerful technique to probe the underlying physics determining gel formation.  An example is Time-Composition Superposition (t-CS) which can be applied to create a mastercurve that describes the class of material and can be used to determine physical characteristics e.g. the cross-link density in polymeric gels \cite{weir2016} and rubbers \cite{ferry} and the strength of interaction between the interface and matrix in dispersions \cite{datta2011}.  The number density of gel-strengthening agents, e.g. cross-links or the filler density, shifts the frequency at which elastic behaviour changes to viscous flow behvaiour, $G'=G''$.  When a dispersed phase interacts strongly with a matrix it is considered an `active filler' and the storage modulus increases with volume fraction \cite{dickinson1999}.  The converse occurs in the case of an inactive filler.  In this paper we measure rheological spectra to understand how the volume fraction affects the gel properties of a colloid-nematic composite.  

A thermotropic nematic liquid crystal (NLC) is composed of rod-like molecules, nematogens, on average oriented along a preferred direction, described by the `director' \textbf{n(r)} \cite{lc}.    The orientation of the director with respect to the direction of flow affects both the viscous response (described by the Miesowicz viscosities)\cite{Beens1983} and elastic moduli. At low frequencies the director rotates in phase with the shear rate and at high frequencies the director rotates in phase with the shear strain, \cite{mather1995} \cite{ternet1999}.  The storage loss $G'$ is elevated at intermediate frequencies due to distortional effects \cite{burghardt1990, mather1996}.    Rey provided a universal theoretical description of the dynamic rheological response of all nematic liquid crystals, valid for a broad range including flow-aligning and tumbling, polymeric and lyotropic nematic materials \cite{rey2006}. Superposition through appropriate scaling of the moduli revealed identical viscoelastic response behaviour for which the loss modulus $G''$ is greater than the storage modulus $G'$ at all frequencies.  The ratio tan${\delta}=G''/G' \sim \omega^{-1}$ at low frequencies before increasing with tan$\delta \sim \omega^{1/2}$  above a resonance frequency $\omega_r$ \cite{rey2006}.  In 1990, Burghardt warned that ``in the presence of monodomains, the estimated time scale for director relaxation becomes short enough that distortional elastic effects may contribute significantly to the macroscopically ob­served viscoelastic response'' \cite{burghardt1990}. 

When a spherical colloid is dispersed in a NLC, nematogens can lie parallel or perpendicular (homeotropic) to the colloid surface\cite{lc,Poulin1997,nehring1971}.  In general terms, the type of defect induced by colloids in NLC depends upon the generalized elasticity $K$, of the nematic phase, the particle radius $r$, and strength of anchoring $W$, of mesogens at the surface of the colloids \cite{lc,Poulin1997}.  Weak homeotropic anchoring with $\frac{W r}{K} \ll 1$ induces the Saturn-ring defect, a defect line (or disclination) that encircles the colloid at an orientation normal to the local director orientation with a topological charge of $s=-1/2$.  Nematogens orient radially from the core of the disclination which is itself isotropic.  Colloids experience highly anisotropic interactions, $U>100 k_B T$, with both attractive and repulsive components depending on the relative orientation of neighbouring colloids and the far-field director \cite{Poulin1997,terentjev1995, fukuda2001,gu2000}; this results in the formation of clusters.  In close proximity, Saturn-ring disclinations can entangle with a range of possible topological configurations \cite{araki2006}.  At sufficient colloid concentrations, $\phi \geq 18 \% $,   `Saturn-ring' disclinations can connect and percolate through a composite thus providing rigidity \cite{wood2011}.  

 We provide new insights on the dynamic behaviour of nematic liquid crystals filled with colloids in the range 20\% $\leq$ $\phi$ $\leq$ 45\%.  Composites were prepared by dispersing poly(methyl methacrylate) spherical particles (PMMA), sterically stabilized by poly-12-hydroxystearic acid (PHSA) in the thermotropic nematic liquid crystal, 4-Cyano-4'pentylbiphenyl (5CB), purchased from Kingston Chemicals (UK) and used as received.  PHSA `hairs' generate homoeotropic anchoring of the nematogens at the colloid surface \cite{wood2011}.  Particle-size was determined on dilution using dynamic light scattering (DLS) and the diameter $d=1.17\pm0.12\mu$m. Colloids were dispersed in hexane and dried thoroughly at $50^{\circ}$C for $>$ 3 days.  The Miesowicz viscosities for 5CB NLC are $\eta_1=110$ mPa.s, $\eta_2=28$ mPa.s and $\eta_3=44$ mPa.s \cite{5cbvisc} and the rotational viscosity $\gamma_r = 87$ mPa.s \cite{rotvisc}.  The splay, twist and bend elastic constants are $k_1=5.9$pN, $k_2=4.5$pN and $k_3=9.9$pN, respectively \cite{5cbelastic}. To make composites, dry PMMA particles (density $\rho_{p} = 1.18$ g/cm$^3$) were added to 5CB (density $\rho_{5CB} \sim 1$ g/cm$^3$) at the appropriate weight fraction and sonicated for 30 mins. Samples were stirred vigorously by hand, with a spatula, prior to placing on a 40 mm stainless steel parallel plate geometry with sand-blasted geometry at a gap of $h=500\pm1\mu$m on a strain-controlled rheometer, ARES G-2 (TA Instruments).  The temperature was maintained at $25\pm1\degree$C.  A pre-shear of 0.1/s was applied before recording oscillatory strain sweeps $0.01\%<\gamma<10\%$.  Angular frequency sweeps $0.1 \geq \omega \geq 600$ rad/s were conducted well above the resonance frequency $\omega_r = 18.65 \frac{k_1}{\eta_1 h^2} < 0.005$ rad/s assuming the relevant lengthscale $h>15 \mu$m. 

\begin{figure}
{\includegraphics[scale=0.4]{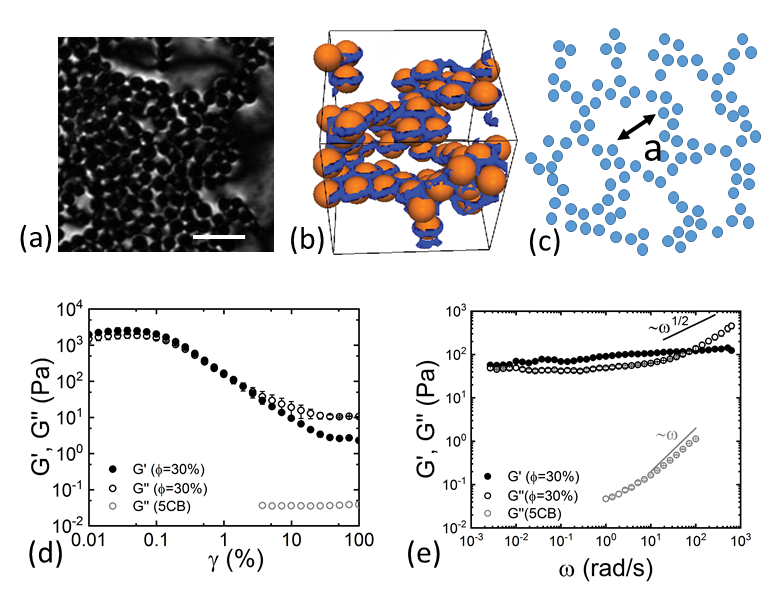}}
\caption{\label{fig1} a) $\phi=0.3$ colloid dispersed in 5CB, sandwiched between two glass slides, and viewed using cross polarised microscopy, scale bar $=5\mu$m. b) Disclinations wrapping around colloids, created through computer simulation \cite{wood2011}. c) Schematic of colloidal structure indicating free nematic domains of size $a$.  Oscillatory rheology measurements on a  $\phi= 30\%$ composite produce d) $G'(\gamma)$ and $G''(\gamma)$ at $\omega =2\pi$rad$/$s and e) $G'(\omega)$ and $G''(\omega)$ at $\gamma=0.6\%$ .  Measurements obtained for pure 5CB are shown in grey.}
\end{figure}

A composite of $\phi=30\%$, flattened between two glass slides, imaged using cross-polarised microscopy, is shown in Fig.\ref{fig1}a.  Isotropic particles appear as black circles on a grey birefringent background generated by the surrounding nematic phase.  Black lines connecting particles are evidence of the disclinations percolating through the composite, predicted through computer simulation, Fig\ref{fig1}b \cite{wood2011}.  In agreement with Kumar \textit(et al.) \cite{Kumar2019}, we observe that colloid free nematic domains are trapped within the colloid structure as indicated by the illustration Fig.\ref{fig1}c. 
Oscillatory rheology measurements, shown in Fig.\ref{fig1}d,  indicate that a $\phi=30\%$ composite is highly elastic with $G' >G''$ in the linear viscoelastic region (LVR) .   The magnitude of the moduli are both $\sim 10^3$Pa, remarkably $5x$ that for the pure liquid crystal, $G'' \sim 10^{-2}$Pa; an increase $<2x$ the background viscosity is expected for active fillers in a polymer network \cite{dickinson1999}.  The LVR extends until a critical strain $\gamma_c = 0.1\%$, beyond which yielding is gradual until $G''$ dominates at $\gamma \sim 1\%$.   Frequency spectra, Fig.\ref{fig1}e, indicate that $G'>G''$ down to the lowest measureable frequencies $\omega \sim$ $0.002$rad/s.   Samples stored in the laboratory have remained stable over a decade suggesting that $G'> G''$ for $\omega <10^{-6}$rad/s.  This indicates that relaxation is not permitted within the system over long time scales.  $G'$ and $G''$ are almost independent of the frequency until a critical frequency $\omega_c$ beyond which $G'' \propto \omega^{1/2}$.  This is unlike the frequency dependence of the pure liquid crystal (shown in grey), for which $G'' \propto \omega$ \cite{rey2006} and other soft systems such as polymer solutions \cite{carrot} and worm-like micelles \cite{zhang}.  

\begin{figure}
 {\includegraphics[scale=0.4]{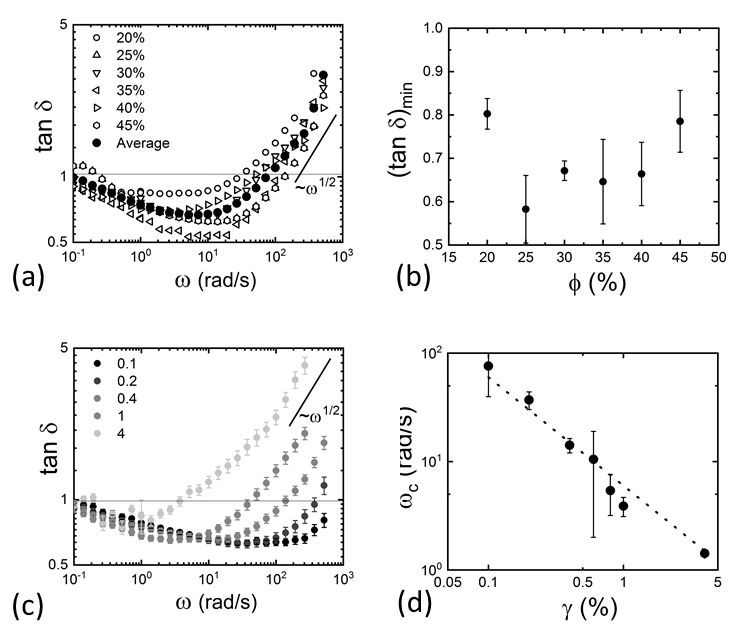}}
\caption{\label{fig2} tan$\delta$ as a function of angular frequency for a) $20\% <\phi < 45\%$ and b) a range of applied strain from $0.1 < \gamma < 4$.  c) The minimum value of tan$\delta (\phi)$ and d) the critical frequency $\omega_c$ at which tan$\delta = 1$.}
\end{figure}

As observed by Wood \textit{et al.} \cite{wood2011} there is a rapid increase in the value of $G'$ for $\phi$ $\geq$ 18\%, consistent with the functional form of $G'_{(LVR)}(\phi) \sim \phi^n$, with n = 2.14$\pm$0.18.  This is consistent with the $G' \sim \phi^2$ dependence observed for colloidal dispersions in an isotropic solvent \cite{mason95}.   Also, $G''\sim \phi^2$ indicating that $G''$ and $G'$ are intimately linked.   tan$\delta$ = $G''/G'$ , shown in Fig.\ref{fig2}a, showed no shift in frequency dependence with filler concentration in contrast to `active' colloid-polymer systems \cite{dickinson1999}. Frequency independence suggests that the nematic phase, solely, determines both the elastic and viscous properties of the composite.  The volume fraction of colloids merely increases the density of disclinations within the system.  This suggests there are no colloid-colloid interactions where $\phi<45\%$.  Experiments were not performed above this limit since samples crumbled and were difficult to handle.  At  $\phi=45\%$ we expect the inter-particle separation $r/d \sim (0.64/0.45)^{1/3} \sim 1.12$ (presuming colloids touch at random close packing $\phi = 64\%$), thus approaching the inter-particle separation, $r/d \sim 1.1$, between colloids entangled by disclinations as predicted by Araki \textit{et. al} \cite{araki2006}.  Insufficient solvent is available to create a disclination dominated composite at higher volume fractions $\phi>45\%$.  
To explore the yielding behaviour, we performed frequency sweeps at higher strains beyond the LVR.  As shown in Fig.\ref{fig2}c a higher applied strain reduces the critical frequency $\omega_c$, at which tan$\delta = 1$ while maintaining the tan$\delta \sim \omega^{1/2}$ behaviour at high frequencies.  In Fig.\ref{fig2}c we find that $\omega_c \sim 1/\gamma$.

To understand the experimental results we now consider the properties of the composite through introducing a simple model. The Frank elasticity of a nematic phase gives rise to disclinations in response to deformation. Each disclination carries a line tension $T=\pi K s^2 ln \frac{L}{r_c}+\pi \sigma_c r_c^2$ \cite{osterman2010}. $r_c$ is the core radius and $\sigma_c$ is the energy density of the disclination, often approximated as $\sigma_c \approx K/r_c^2$ \cite{osterman2010}.  $L$ is the linear dimension of the region of director deformation around the disclination.   For a thermotropic liquid crystal $T\sim 100$pN where $r_c \sim 5$nm and $L \gg r_c$.   The microstructure of our composite is a network of colloidal clusters, bound together by disclinations and pure nematic regions, of size $a$, as illustrated in Fig.\ref{fig1}c.   

The storage modulus describes the elasticity of the material and is the sum of two different contributions $G'=G'_f+G'_d$.  $G'_f$ describes the elasticity of a nematic liquid crystal flowing between parallel plates.  It was described by Rey \cite{rey2006} and at frequencies $>\omega_r$, $G'_R=\left(\frac{\alpha_2}{\gamma_r})^2(2 \omega \eta_{bend}\right)^{1/2}$.  For 5CB $G'_f \sim 10^{-5}$Pa.

The elasticity of a liquid crystal, caused by the presence of disclinations, at an average separation of $d_{net}$, was described by Weitz \textit{et al.} as   $G'_d=\left(\frac{T}{d_{net}}\right)^2$ \cite{Weitz}.   Since experiments indicate that $G'\sim\phi^2$ we use $G'_d=\left(\frac{T}{d_{net}}\right)p(\phi)^2$ where $p$ is an unknown factor.     In the colloid-rich regions, disclinations are separated by $d_0= 1.1 D$ where $D$ is the diameter of the particles.  We assert that disclinations yield at a critical strain of $\gamma_c=\frac{r_c}{1.1D}$ where $r_c$ is the radius of a disclination.  Above this strain, disclinations yield such that the density of the disclinations increase as $d_{net}=d_0 \gamma_n$ where $\gamma_n=\left(\frac{\gamma_c+\gamma}{\gamma_c}\right)$.

The viscous behaviour of the composite is determined by two contributions $G''=G''_f+G''_a$.   We assume the term, $G''_f = \eta_2 \omega$ describes the nematic flow behaviour at high shear where $\eta_2$ is the Miesovicz viscosity associated with the director aligning along the shear direction.  $G''_a$ describes the viscous contribution within confined regions.  The Ericksen number describes the competition between flow-induced and boundary-induced orientation within a nematic monodomain $E_r=\frac{L^2 \gamma_r \dot\gamma}{K}$ where $\gamma_r$ is the rotational viscosity, $\dot\gamma$ is the shear rate and $L$ is the relevant lengthscale.  Reorientation of the director can occur when $E_r >1$.  For polydomains in liquid crystal polymers it has been argued that disclinations can act as an internal wall for the bulk nematic and that the director can become anchored at these internal walls \cite{acierne}.  In a confined region, the viscosity increases as $\eta_a \sim \gamma_r Er^{-1/2}$ \cite{rheo}.  We assume $G''=p(\phi)^2 \eta_a \omega_n$    In oscillatory flow, $E_r= \frac{a^2 \gamma_r \gamma_n \omega_n}{2 \pi K}$ where $a$ is the average dimension of colloid-free nematic domains within the composite, the size of which will be determined by the ratio of the elasticity, $K$, to the anchoring energy, $W$ of nematogens at the surface of colloids such that $a \sim K/W$.  We presume $W=1.5X10^{-7}$ J/m, consistent with earlier estimations \cite{terentjev1995} and $K=5.5$pN.  Nematic domain growth forces colloids to approach one another until $a \sim 10\mu$m, unless colloid crowding limits growth.  The frequency $\omega_n = \omega_c + \omega$ where $\omega_c = \frac{2 \pi K}{a^2 \gamma_r \gamma_n}$ is the critical frequency at which flow overcomes the elasticity of the composite.     Inserting our definition of the Ericksen we find that 
\begin{equation}\label{xx}
G''_a=p(\phi)^2\left[\left( \frac{2 \pi K}{a^2 \gamma_n}\right) + \left(\frac{2 \pi K \gamma_r \omega}{a^2 \gamma_n}\right)^{-1/2}\right]
\end{equation}
The first term describes the plateau $G''_d$ which is independent of frequency and dominates at low frequencies.  The second term, $G''_c \propto \omega^{1/2}$ overcomes $G''$ beyond the critical frequency, $\omega_c$. 

\begin{figure}
  \centering
  {\includegraphics[scale=0.4]{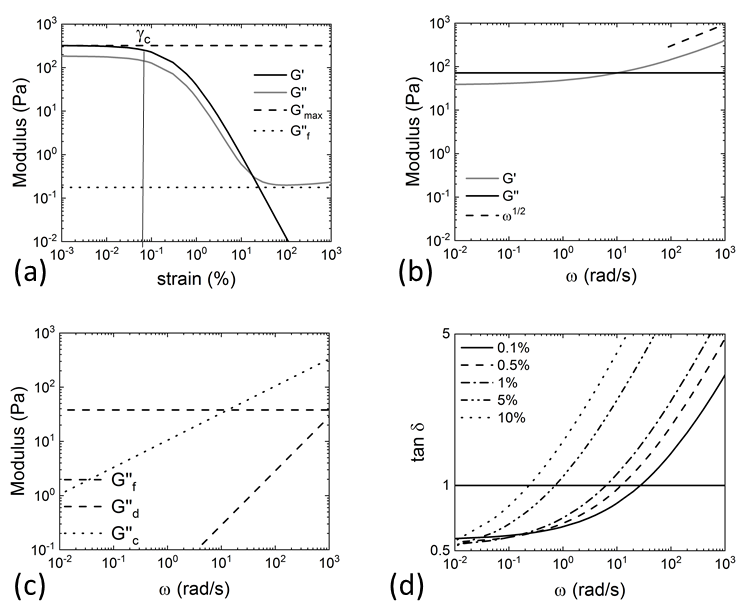}}
\caption{\label{fig3} a) Theoretical prediction of $G'(\gamma)$ and $G'(\gamma)$ for $\omega = 2\pi$rad$/$s and b) $G'(\omega)$ and $G'(\omega)$ for $\gamma=0.6\%$. c) Components $G''_f$, $G''_d$, $G''_c$  contributing to the total loss modulus $G''$ shown in b).  d) Plot of tan$\delta (\omega)$ for  $0.1\% < \gamma < 10\%$}
\end{figure}

The theoretical predictions are presented in Fig.\ref{fig3}.  $G'$ and $G''$ show strong similarities to the experiment results with $G'>G''$ and low frequencies and $G'' \propto \omega^{1/2}$ on yielding.     The components of $G''_a=G''_d+G''_c$ and $G''_f$ are compared in Fig.\ref{fig3}c.  It is clear that for confined regions, $G''_d > G''_f$ thus suppressing the $G'' \propto \omega$ dependence observed in similar experiments for low colloid fractions \cite{Weitz}. At low frequencies  $tan \delta(\omega \to 0) = \frac{2 \pi K}{T} \frac{1}{\gamma_n}\frac{d_{net}^2}{a^2}$  which indicates that tan$\delta (\omega \to 0)$ is dependent on the ratio of lengthscales associated with the disclination sepration and the nematic domain size.  Non-monotonic behaviour in the range $0.6 <$(tan$\delta)_{min}< 0.8$, was observed experimentally, shown in Fig.\ref{fig2}b.    Therefore, we might expect that tan$\delta$ could be modulated through the colloid size, elasticity and the anchoring strength of nematogens.  In our model, we used $p(\phi)=6$; we invite computer simulators to clarify the meaning and value of $p(\phi)$.

In conclusion, we bring experiments and theory together to understand the dynamic behaviour of concentrated colloidal dispersions in a nematic phase with nematogens oriented homeotropically at the colloid surface with weak anchoring conditions $Wa<K$.  This class of glassy soft-solid \cite{glass} has received scant attention to date.  We find the dispersed phase is an `active filler', increasing $|G| \sim \phi^2$.  Unlike other `active fillers systems, the frequency behaviour is independent of the volume fraction.  The magnitude of $|G|$ and is sensitive only to the disclination density and colloid-free nematic domain size.    Beyond the linear viscoelastic region, disclinations break, and the material yields at a critical frequency which is inversely proportional to the strain applied.   After yielding, the viscous behaviour $G'' \propto \omega^{1/2}$.  Similar behaviour has been observed previously but not explained \cite{Kumar2019,Yamamoto}.  This is the first time that the dynamic behaviour has been studied systematically and the stability and yielding behaviour has been attributed to the Ericksen number and elevated viscosity within confined nematic domains, as described by Eq. (1).  .  Our work provides guidelines for developing new composites with superior physical stability through dispersing colloids in nematic liquid crystals.

We thank A. B. Schofield for synthesizing PMMA particles. This work was funded by a Royal Society Industry Fellowship and a Sydney-Andrew Scholarship funded by Society of Chemical Industry (SCI).

\end{document}